\documentclass[conference]{IEEEtran}
\IEEEoverridecommandlockouts
\usepackage{cite}
\usepackage{amsmath,amssymb,amsfonts}
\usepackage{algorithmic}
\usepackage{graphicx}
\usepackage{amsmath}
\usepackage{textcomp}
\usepackage{mathrsfs}
\usepackage{amsmath}
\usepackage{mathtools}

\usepackage{color}
\usepackage{multirow}
\usepackage{xcolor}
\usepackage{subfigure}
\usepackage{colortbl}
\definecolor{mygray}{gray}{.9}
\usepackage{threeparttable}
\usepackage{color, soul}
\usepackage{ulem}
\usepackage[justification=centering]{caption}
\usepackage[numbers,sort&compress]{natbib}
\usepackage{xcolor}
\usepackage{breqn}
\usepackage{tikz}
\usepackage{textcomp}
\usepackage{lipsum}

\newcommand{\mv}[1]{{\color{black}#1}}

\newcommand\copyrighttext{%
  \footnotesize \textcopyright 2024 IEEE. Personal use of this material is permitted.
  Permission from IEEE must be obtained for all other uses, in any current or future
  media, including reprinting/republishing this material for advertising or promotional
  purposes, creating new collective works, for resale or redistribution to servers or
  lists, or reuse of any copyrighted component of this work in other works.
  DOI: 10.1109/ISQED57927.2023.10129330.}
  
\newcommand\copyrightnotice{%
\begin{tikzpicture}[remember picture,overlay]
\node[anchor=south,yshift=10pt] at (current page.south) {\fbox{\parbox{\dimexpr\textwidth-\fboxsep-\fboxrule\relax}{\copyrighttext}}};
\end{tikzpicture}%
}

\makeatletter
\def\thickhline{%
  \noalign{\ifnum0=`}\fi\hrule \@height \thickarrayrulewidth \futurelet
   \reserved@a\@xthickhline}
\def\@xthickhline{\ifx\reserved@a\thickhline
               \vskip\doublerulesep
               \vskip-\thickarrayrulewidth
             \fi
      \ifnum0=`{\fi}}
\makeatother

\newlength{\thickarrayrulewidth}
\def\BibTeX{{\rm B\kern-.05em{\sc i\kern-.025em b}\kern-.08em
    T\kern-.1667em\lower.7ex\hbox{E}\kern-.125emX}}
    
\begin{document}

\title{CMDS: Cross-layer Dataflow Optimization for DNN Accelerators Exploiting Multi-bank Memories}


\author{
    \IEEEauthorblockN{Man Shi\IEEEauthorrefmark{1}, Steven Colleman\IEEEauthorrefmark{1}, Charlotte VanDeMieroop\IEEEauthorrefmark{1}, Antony Joseph\IEEEauthorrefmark{2}, \\
    Maurice Meijer\IEEEauthorrefmark{2}, 
    Wim Dehaene\IEEEauthorrefmark{1}, Marian Verhelst\IEEEauthorrefmark{1}} 
    \IEEEauthorblockA{\IEEEauthorrefmark{1}MICAS-ESAT, KU Leuven,
    \IEEEauthorrefmark{2}NXP Semiconductor}
    \IEEEauthorblockA{ \{man.shi, steven.colleman, wim.dehaene, marian.verhelst\}@kuleuven.be, \{antony.joseph, maurice.meijer\}@nxp.com}
}
\maketitle
\copyrightnotice
\begin{abstract}
Deep neural networks (DNN) use a wide range of network topologies to achieve high accuracy within diverse applications. This model diversity makes it impossible to identify a single "dataflow" (execution schedule) to perform optimally across all possible layers and network topologies. Several frameworks support the exploration of the best dataflow for a given DNN layer and hardware. However, switching the dataflow from one layer to the next layer within one DNN model can result in hardware inefficiencies stemming from memory data layout mismatch among the layers. Unfortunately, all existing frameworks treat each layer independently and typically model memories as black boxes (one large monolithic wide memory), which ignores the data layout and can not deal with the data layout dependencies of sequential layers. These frameworks are not capable of doing dataflow cross-layer optimization. 
This work, hence, aims at cross-layer dataflow optimization, taking the data dependency and data layout reshuffling overheads among layers into account. Additionally, we propose to exploit the multi-bank memories typically present in modern DNN accelerators towards efficiently reshuffling data to support more dataflow at low overhead. These innovations are supported through the \uline{\textbf{C}}ross-layer \uline{\textbf{M}}emory-aware \uline{\textbf{D}}ataflow \uline{\textbf{S}}cheduler (CMDS). CMDS can model DNN execution energy/latency while considering the different data layout requirements due to the varied optimal dataflow of layers. 
Compared with the state-of-the-art (SOTA), which performs layer-optimized memory-unaware scheduling, CMDS achieves up to $5.5\times$ energy reduction and $1.35\times$ latency reduction with negligible hardware cost.

\end{abstract}


\section{Introduction and motivation}
\IEEEPARstart{D}{eep} neural networks have become a fundamental building block of massive smart sensing applications due to their superior accuracy in various image and audio processing tasks. However, the accuracy boost from DNNs comes at the cost of high computational complexity and diversity of network topologies. Some DNN models, such as ResNet \cite{resnet}, MobileNetv2 \cite{m2}, and YamNet \cite{ya}, are used in image and audio classification tasks. The layers constituting these networks vary widely in terms of layer types (classical convolutional layer, point-wise layer, depth-wise layer, fully connected layer) and layer shapes (widely differing in height, width, input channel, and output channel). The dissimilarity of DNN layers makes it infeasible to specify a fixed dataflow (execution schedule) for machine learning (ML) hardware accelerators to perform optimally across all possible layers and networks \cite{sch}. Several studies \cite{aicas, eyeriss, hnpu} have shown the importance of supporting multiple dataflows within a single DNN accelerator, with demonstrated gains of up to 60\% lower network inference EDP (energy-delay-product) compared to a single dataflow \cite{aicas}. These benefits resulted in many design space exploration frameworks (such as ZigZag, Timeloop, Maestro \cite{zigzag, timeloop, maestro}) that can systematically explore the optimal dataflow for mapping a specific neural network (NN) layer on a specific DNN accelerator. 

\begin{figure}[tb]
\centering 
\vspace{-0.1cm}
\includegraphics[width=1\columnwidth]{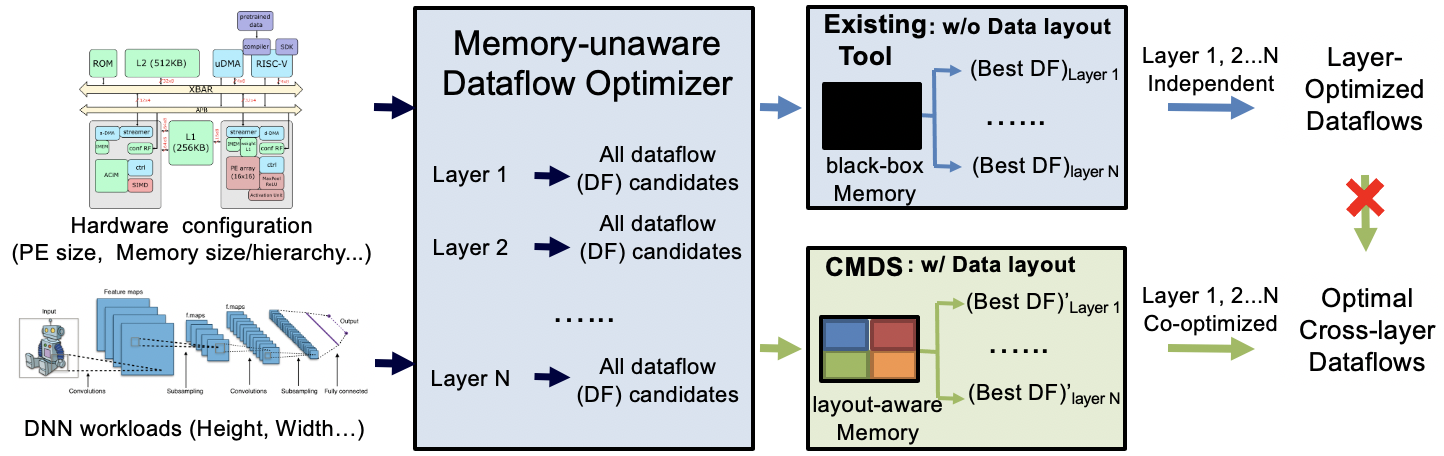}
\caption{Memory-unaware dataflow (DF) optimizer v.s. CMDS}
\vspace{-0.4cm}
\label{fig:flow} 
\end{figure}

Nevertheless, all the existing dataflow optimizers treat each NN layer as an independent workload, ignoring the data dependency of sequential layers, and fail to access the overheads of supporting multiple dataflow within a single accelerator. As shown in Fig. \ref{fig:flow} (in blue), these existing tools naively select the minimal energy and/or latency dataflow for each layer individually on the targeted hardware, and subsequently accumulate the results \cite{timeloop} to get the complete NN evaluation results. In real scenarios, the layer-independent optimization results in different optimal dataflows among the layers and causes distinct data layouts in memory.
The consequences of the mismatched data layouts for sequential layers are not modeled in the SOTA frameworks \cite{zigzag, timeloop, maestro}. Because they model the data memories as black boxes,  which can continuously read/write the same amount of data per cycle (its memory port width), regardless of its internal data organization. This mechanism is especially problematic for activation memories, whose data are generated on-chip and can not be arranged in arbitrary order. In practice, the data layout mismatch between layers will under-utilize the memory port width, which can significantly impact DNN acceleration efficiency. As such, the "best" dataflows for each layer searched by existing memory-unaware tools can not achieve the optimal end-to-end NN inference.

To overcome this hurdle, we propose a \uline{\textbf{C}}ross-layer \uline{\textbf{M}}emory-aware \uline{\textbf{D}}ataflow \uline{\textbf{S}}cheduler (CMDS),
which is capable of modeling the memory's internal data layout to achieve effective cross-layer dataflow optimization (Fig.~1). The contributions and innovations of this work are summarized as follows:
\begin{itemize}
\item Proposes a concise way to describe the memory structure and exploit multi-bank memories for efficient data reorganization to reduce the cost of supporting layer-to-layer dataflow flexibility in NN execution (Section III).%
\item Presents CMDS that can perform cross-layer dataflow optimization taking into account the memory data layout and resulting energy/latency consequences, offering significant efficiency savings over the SOTA (Section IV). 
\item Integrates CMDS with SOTA layer-wise dataflow optimizers, using a wrapper modeling the impact of memory data layout and an aligned cost model for data reshaping towards complete cross-layer scheduling. Experimental results are conducted in combination with the layer-wise evaluation tool ZigZag \cite{zigzag}, showing up to $5.5\times$ energy reduction and up to 1.35× less latency. (Section V).
\end{itemize}



\section{Background and SOTA} 
A typical NN layer can be represented as a set of nested for-loops (in Fig.~\ref{fig:sutu}), where OX (IX), OY (IY), K (C), and B indicate the dimensions of the output (input) feature map; and FX, FY, C represent the shape of the weight kernels. These for-loops can be tiled and ordered to many smaller nested for-loops. Therefore, the NN layer can be processed in hardware with
a large variety of potential dataflows, each characterized by a different parallelization (spatial unrolling, SU) and loop tiling/ordering strategies (temporal unrolling, TU).  
\subsection{Spatial and Temporal Unrolling}
\begin{figure}[tb]
\centering
\includegraphics[width=0.85\linewidth]{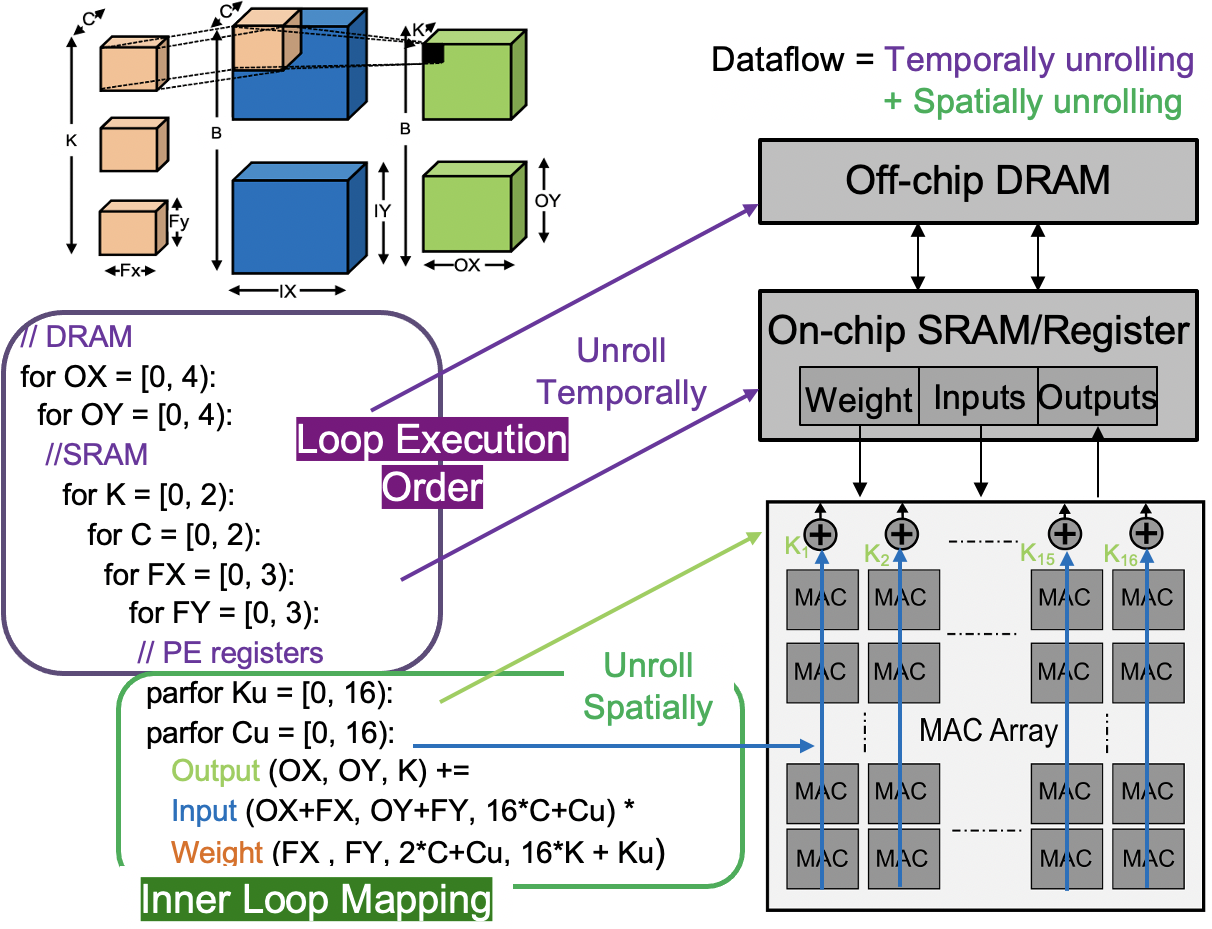}
\caption{A layer with $C = K = 32$, $OX = OY = 4$ and $FX = FY = 3$ is evaluated on a PE array of size 16x16} 
\label{fig:sutu}
\vspace{-0.4cm}   
\end{figure}

In Fig. \ref{fig:sutu}, an example is used to show one possibility of how a layer is mapped on a typical accelerator consisting of a PE array of size 16x16 and a memory hierarchy.

\textbf{Spatial unrolling:} The SU deals with the spatial parallelization of the layer's loop dimensions within one clock cycle. In other words: how the most inner loop dimensions are unrolled upon the 2D MAC array (Fig. \ref{fig:sutu} bottom). This paper will denote this parallelism with the suffix '\textbf{u}'. Specifically, $OX_u, OY_u, C_u, K_u, FX_u, FY_u$, indicate the number of data elements required in each cycle from each loop dimension. 

\textbf{Temporal unrolling:} The remaining loop dimensions
are unrolled temporally (Fig.~\ref{fig:sutu} top). The ordering/tiling schemes of loops impact the data allocation in the memory hierarchy when processing different NN layers.


\subsection{Data formatting layout}
Supporting multiple TUs within a single accelerator typically can be realized through software or FSM control. 
In contrast, the support for multiple SUs within one DNN accelerator comes at a hardware cost. As declared in \cite{eyeriss}, a reconfigurable network-on-chip (NOC) makes it feasible to support flexible data fetching for the PE array at less than $3\%$ of the total area cost and $6\%-10\%$ of the total energy consumption. However, changing SUs does not only come with the small cost of a flexible NOC to distribute the data to the PE array, but also results in ineffectual memory accesses caused by a layout mismatch between the data stored in memory and the data needed by the PE array. The layout of the features data can be completely different for various layers if their SUs are different. We have observed that the data layout mismatch between two data-dependent layers eliminates the efficiency of supporting multiple dataflows. Considering the memory data layout is crucial for reaching optimal end-to-end NN inference. 


Therefore, recent DNN processors, such as Diana \cite{diana}, utilize a reshuffling buffer to reformat features between layers to ensure that all data required by the PE array within a single clock cycle is grouped within one wide memory word. However, the main drawback of this solution is the enormous area cost and high design complexity of such a flexible reshuffling buffer \cite{aicas}. 
Reshuffling over DRAM \cite{dram} does not solve this problem, as costly external memory accesses undo the benefits of combining different SUs. As a result, several SOTA accelerator designs only support a single SU or a limited set of similar SUs \cite{evolver, cpu}. However, these similar SUs are typically too narrow to efficiently cover NN layers' diversity. To tackle this problem, we propose a novel scheme that exploits existing multi-bank memories to perform efficient data reshuffling. 

\section {Data reshuffle through multi-bank memories}
DNN accelerators heavily exploit on-chip memory (mainly SRAM) to store activations and weights, avoiding costly external memory accesses. These large on-chip memories are typically implemented with many memory banks, where each bank is independently addressable. For efficiency reasons, each bank can not be made too large or too fine-grained. However, their granularity can still be exploited for efficient data reshuffling. 

\begin{figure}[tb]
\centering
\includegraphics[width=0.95\linewidth]{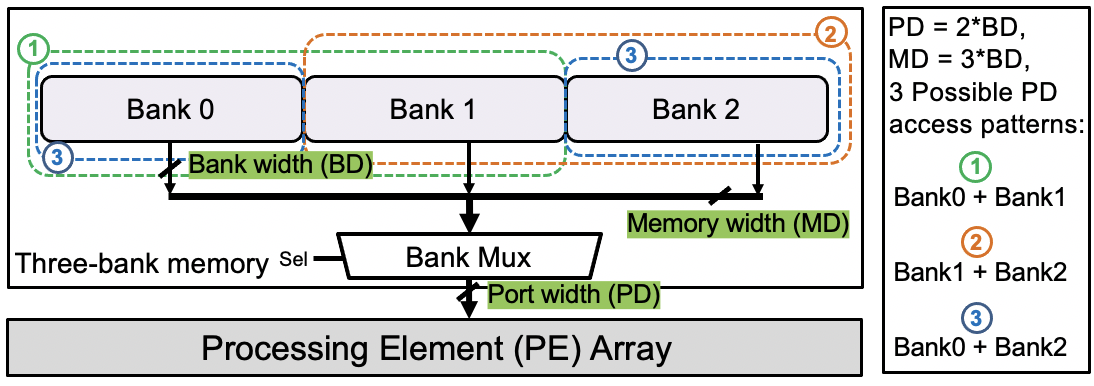}
\caption{Concept of memory BD, PD, and MD visualization.}
\label{fig:3_comb}
\vspace{-0.4cm}
\end{figure}
 
To this end, we introduce the concepts of Bank width, Port width, and Memory width (illustrated in Fig.~3), which are used to represent the memory layout in this work concisely. 
We denote the row dimension of a bank by the \textbf{\uline{B}ank wi\uline{d}th} (BD). The maximum number of banks that can be activated in parallel impacts the memories \textbf{\uline{P}ort wi\uline{d}th} (PD), making a PD multiple of BD. Note that within one cycle, the port can only access separate memory banks. If data from different rows of the same bank is required in parallel, the system has to stall. This phenomenon is called bank contention \cite{sram}. It is possible to increase the number of memory banks beyond $PD/BD$ to alleviate this phenomenon, which increases the effective total \textbf{\uline{M}emory wi\uline{d}th} (MD), computed as the total number of banks $\times$ BD. The larger effective MD allows reducing data conflicts by smartly distributing access to the handled data across more banks. In Fig.~\ref{fig:3_comb}, PD comprises two of these three banks. Thus $PD = 2 \times BD; MD = 3 \times BD$. Larger MD hence raises the flexibility of what data can be read in parallel over the memory port, here having three possible bank access patterns. CMDS will later exploit this multi-bank structure of memory to reorder data. Compared with reshuffling buffer, reshaping data realized by this multi-bank structure can avoid massive extra area cost, which allows CMDS to be able to efficiently support various DNN dataflows with minimal hardware overhead.


Note that the optimal memory banking design solution depends on the hardware resource budgets. This paper only utilizes the existing multi-bank memory structures in accelerators to do intelligent scheduling and does not cover the optimal design solution for memory banking.

\section {Cross-layer Memory-aware Dataflow Scheduler} 
In this section, the CMDS flow is introduced. To explore and optimize the cross-layer dataflow, two essential modifications are required : 1) extend the input hardware configuration of scheduling frameworks with the BD and MD parameters (PD is typically already a configuration), providing a more accurate memory modeling; 
2.) enhance the scheduler to exploit data layout options within the BD/PD/MD constraints. 
To decrease the complexity, two reasonable assumptions are adopted: 
\begin{itemize}
     \item All data words, BD, PD, and MD, are a power of two. 
     \item The number of PEs is a power of two, triggering all unrolling factors of SU and TU to be a power of two.
\end{itemize}

\begin{figure*}[tb]
\centering 
\includegraphics[width=1.87\columnwidth]{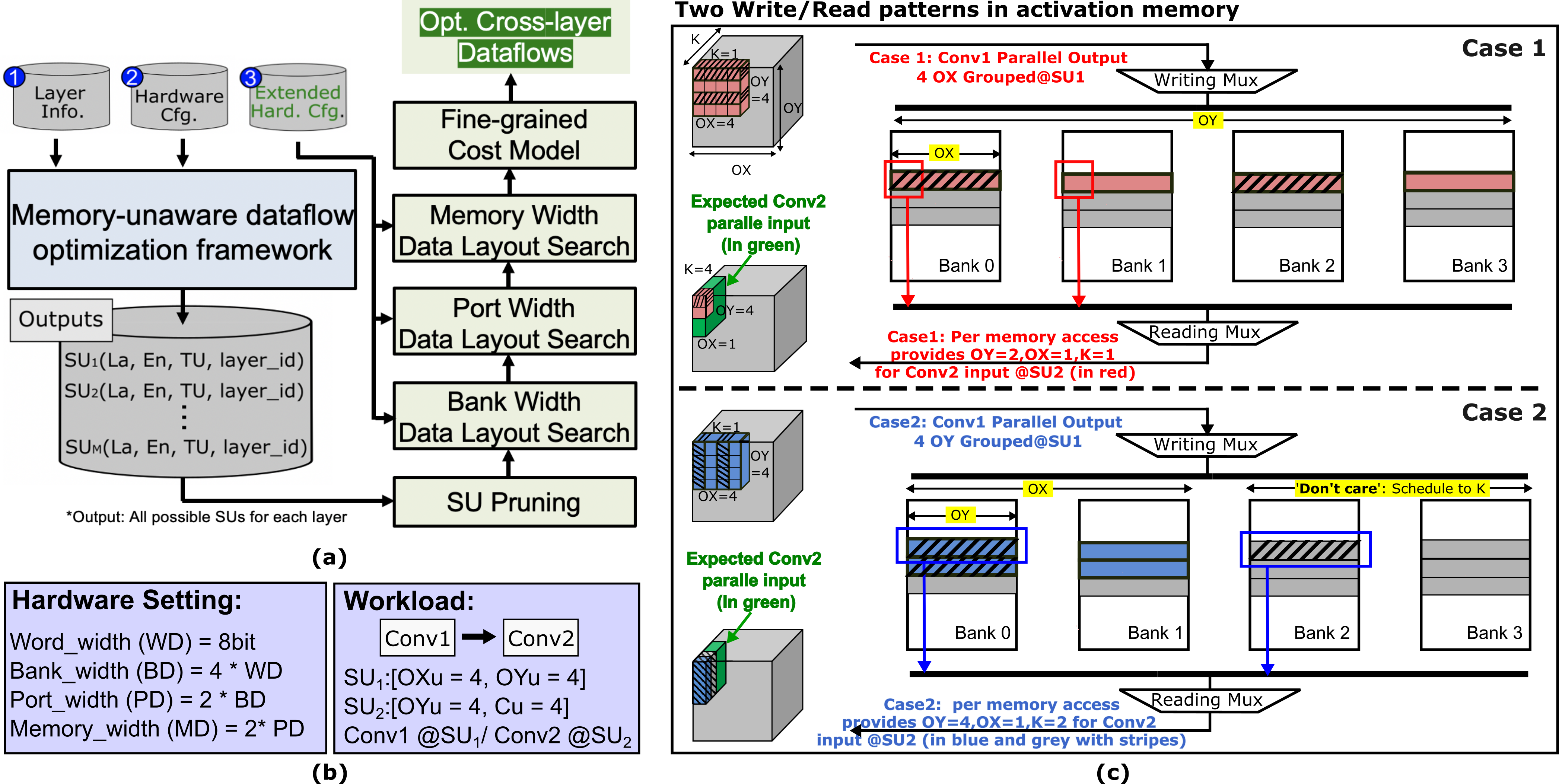}
\caption{(a) CMDS automation flow. (b) Hardware setting and example workload; (c) CMDS BD/PD/MD scheduling example}
\label{fig:main} 
\vspace{-0.5cm}
\end{figure*}

Fig.~\ref{fig:main}(a) shows the full CMDS tool flow: The flow first calls any SOTA layer-wise optimizer (such as ZigZag, Timeloop, and so on)
to derive for each layer the optimal TU and its resulting energy/latency for all SUs supported by the accelerator under study. 
Secondly, the tool performs a pruning step to shrink the SU pool for each layer to speed up the exploration by pruning away sub-optimal SUs. 
Next, CMDS utilizes the extended hardware configuration to sequentially explore the possible data layout for each NN layer in the function of BD/PD/MD, taking into account the pruned SU pool of preceding and succeeding layers. Finally, a fine-grained cost model is utilized to evaluate cross-layer dependencies and the PD/MD/BD effects (introduce in section V).
This allows CMDS to derive the cross-layer optimized dataflows and memory layout eventually. The following sections will cover each step in detail.
\vspace{-0.1cm}

\subsection{SU pruning}
A small experiment running ZigZag 
For a single Conv. layer (second layer) of ResNet18, the SOTA layer-wise optimizer ZigZag identifies 9960 feasible SUs.
Such a large number of mapping possibilities per layer outcomes a prohibitively large number of SU combinations across all layers of the NN. To this end, an
\textit{SU pruning step} for each layer prunes away these SUs, whose performance is significantly worse than the best SU (for single independent layer) performance.
The pruning metric $P$ can be defined as latency, energy, or EDP. The harshness of pruning is controlled with a threshold $\theta$:
\begin{equation}
 \frac{P_{SU} - P_{SU_{min}}}{P_{ideal\_network}} \leq \theta ; ~ P_{SU} \in [P_{SU_{min}},..., P_{SU_{max}}],
\label{eq1}
\end{equation}
in which $P_{SU}$ is the performance of executing the current layer with the spatial unrolling $SU$, $P_{SU_{min}}$ is the ideal performance achieved by the most optimal SU for the current layer. $P_{SU}$ and $P_{SU_{min}}$ are the immediate outputs from ZigZag without data layout awareness. This performance degradation is normalized to the ideal NN performance of all layers together $P_{ideal\_ network}$ (derived by adding up the respective $P_{SU_{min}}$ of each layer in the NN model), to give more freedom to the SU of non-dominant layers. In this way, dominant layers can keep optimal SU candidates by adjusting the SU of non-dominant layers to match the data layout, as non-dominant layers have more SU alternatives. 
Practically, this pruning step retains for each layer these SUs, which do not degrade the overall NN performance beyond a tolerable limit $\theta$. It is vital to employ a proper $\theta$: the pruning step will prune away massive SUs for each layer if the $\theta$ is too small, even though some SUs have negligible performance degradation. The resulting tiny SU pool that is retained would, however, provide only few SU alternatives for each layer, which prevents finding the best SU combination for the whole NN with data layout awareness.
In contrast, the search time is too long to efficiently explore the best SU combination under the data layout consideration if the $\theta$ is too large. In our experiments, we observe a good balance between scheduling flexibility for data layout optimization and exploration cost when $\theta$ is set to $0.1$. This SU pruning enables a search time reduction for BD/PD/MD data layout optimization with a factor of more than 1000.

\vspace{-0.1cm}
\subsection{Data layout search within bank width}

Next, the best data layout will be searched for each layer's output, given the retained SU pool per layer. The investigation starts by optimizing the data layout enclosed within one row of a memory bank, namely the bank width, BD.
Fig.~\ref{fig:main}(c) depicts the importance of this BD data layout with an example using 8-bit data words, four words making up one BD. 

As the SU of Conv1, $SU_1$, generates 4x4 final outputs within the same clock cycle along the $OX|OY$ dimensions, multiple BD layouts are possible. 
In Case1 of Fig.~\ref{fig:main}(c), the outputs are grouped along the OX dimension, storing 4 OX adjacent words in 1 memory bank row, populating different rows for different OY entries. When the next layer, Conv2, uses $SU_2$, it needs 4x4 inputs in parallel along the dimensions $OY|K$ of Conv1 (Conv2's C equals Conv1's K). 
However, this expected data parallelism of $SU_2$ conflicts with the OX-grouped BD data layout of the $SU_1$, output, as it can utilize only a single data word from each fetched bank row, visualized in Case1. This conflict will cause several issues: 1.) The core can not fully utilize the port width to access the required input data, resulting in latency stalls; 2.) The energy to access an incomplete BD is close to the full BD access cost, resulting in 4x higher energy for total activation data accesses. Alternative BD data layouts could resolve this. In Case2 of Fig.~\ref{fig:main}(c). Conv1 outputs can alternatively be packed along the OY dimension. This BD data layout is compatible with the outputs generated by $SU_1$ as well as the input requirements of $SU_2$. As such, all data accesses can happen at BD-granularity, avoiding wasted memory accesses. This example proves that the BD data layout is of utmost importance, as improper BD data layouts can potentially undo all benefits of combining different optimal SU for individual layers.     

To efficiently explore the optimal BD data layout, CMDS assesses the similarities of SUs. It searches for shared SU unrolling factors among these retained SUs, preventing partial BD data accesses. For instance, the common unrolling factor of $SU_1$ and $SU_2$ in the example is `$OY_u=4$', which is both generated in parallel and consumed in parallel, making it wise to group data together along the OY dimension. CMDS aims to find a common BD data layout for all layers in the NN. To this end, CMDS first constructs all $OX|OY|K$ combinations which fit the width of the one bank row, BD. 
CMDS assesses each BD candidate to check whether it is compatible with at least one retained SU of each layer in the NN and considers the BD layout `valid' when this is the case. A set of SUs with BD-sized common unrolling factors is retained for each valid BD data layout. Finally, the SU pool per layer is pruned to retain all SUs that include these valid BD layouts for the rest of the exploration for the PD/MD data layout.

\subsection{Data layout search within port width }
The PD data layout is optimized in this subsection.
In the example of Fig.~\ref{fig:main}(c), $PD=2\times BD$, i.e. two banks can be read or written in parallel. The PD data layout relates to what spatial factors should be unrolled along the memory port. The flexibility of the multi-bank memory allows different spatial unrolling parameters for write and read accesses, denoted by the \textbf{\uline{W}rite PD} (WPD) and \textbf{\uline{R}ead PD} (RPD), respectively. Case2 presents one possibility of WPD data layout 
based on $SU_1$: beyond 4-OY,
4-OX parallelism can be additionally unrolled either spatially or temporally. Case2 unrolls 2-OX spatially, and the remaining 2-OX temporally 
across two memory write cycles, 
defining the WPD data layout as $[OY_u=4, OX_u=2]$. Similarly, the RPD data layout of Conv2 needs to contain the expected inputs of $SU_2$. 
It is straightforward that the WPD data layout is determined by the SU of a specific layer ($layer_i$), the RPD data layout is determined by the SU of a layer that consumes the outputs of $layer_i$ as inputs. In practice, the optimal PD data layout must contain the valid BD layout (searched from IV-B) as part of unrolling factors to fully use the memory port width. Otherwise, mismatched PD and BD data layouts will cause abundant wasted memory accesses. Note that the RPD and WPD can differ and change from layer to layer, as long as the combination is supported by the MD layout optimization of the following subsection.

\subsection{Data layout search within memory width}
Finally, CMDS optimizes how data is arranged across the entire memory width (MD). Smart MD data layout management will not only diminish the number of stall cycles, but will later also be leveraged
to implement low-cost data reorganization between subsequent layers for full utilization of the WPD/RPD. Going back to our example of Fig.~\ref{fig:main}(c), Case2, the WPD data layout was defined as $[OY=4, OX=2]$. This WPD data layout allows us to, e.g., transfer data to bank0 and bank1 in the first write cycle. In the second cycle, we have multiple options: write the remaining data to bank2 and bank3 (as visualized in Case1) or write during the next cycle into the next row of bank0 and bank1, as depicted in Case2, creating an alternative MD data layout. 
The MD hence offers the flexibility to optimize the temporal access scheme 
to be compatible with the expected RPD data layout of the next layer, ensuring the required data can be fetched in as few cycles as possible. In our example, Case2 optimally unrolls 2-K parallelism along the MD dimension, resulting in the MD layout of $[OY=4, OX=2, K=2]$, which supports both previous layer's WPD and next layer's RPD data needs to avoid extra memory stalls. 
 \begin{figure}[tb]
\centering 
\includegraphics[width=1\columnwidth]{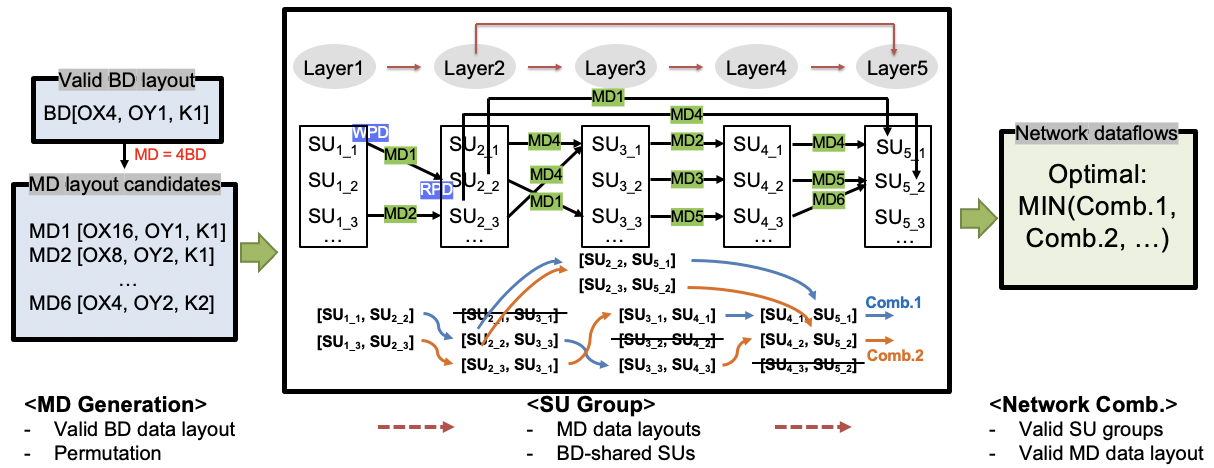}
\caption{Cross-layer dataflow exploration for 5-layer NN}
\label{fig:flow2} 
\vspace{-0.3cm}
\end{figure} 
\subsection{Bringing it all together}
To simplify the explanation, Fig.~\ref{fig:main}(c) is depicted assuming a simple concatenation of two convolutional layers. The realistic model can have much more complex layer dependencies, though. To ensure accurate modeling, CMDS also supports layers with incoming skip connections \cite{dp}, such as residual connections in ResNet. CMDS can explore the data layout of WPD/RPD/MD for multiple data-dependent layers simultaneously based on the analysis above. 
Fig.~\ref{fig:flow2} presents how CMDS explores the best dataflows for the whole NN (e.g., five layers NN in which layer3 and layer5 both receive the outputs of layer2 as inputs).
As BD is a subset of PD/MD, CMDS first constructs all potential MD data layout candidates for each valid BD data layout (from section IV-B), constrained by the total number of banks: $\frac{MD}{BD}$. Next, the allowed WPD and RPD data layouts can be determined from the SUs of any two data-dependent layers (from section IV-C). CMDS accesses the retained SU pool and explores if an MD layout candidate simultaneously contains the WPD layout for a specific SU of $layer_i$ and the RPD layout for the SUs of all $layer_i$'s data-dependent layers. These SUs will be grouped if such an MD candidate exists. 

Specifically, in the example of Fig.~\ref{fig:flow2}, layer2's outputs stream into layer3 and layer5 as inputs data. This means the output data 
layout of layer2 has to be compatible with both layer3 and layer5's input reading pattern: the first MD layout candidate (MD1) presented in Fig.~\ref{fig:flow2} 
includes WPD layout of $SU_{2\_2}$ ($SU_{x\_y}$ means the No. y SU of layer x), RPD layout of $SU_{3\_3}$ and RPD layout of $SU_{5\_1}$ simultaneously, hence pairing these SUs based on data dependency: as $[SU_{2\_2}, SU_{3\_3}]$, $[SU_{2\_2}, SU_{5\_1}]$, respectively.  
Sets of SU groups are generated between any two data-dependent layers. CMDS connects all successive grouped SU pairs with a common SU to create the construct the list of all feasible NN dataflow combinations. For each NN dataflow combination, the total NN energy and latency are estimated, taking all memory-related stalls into account (Section \ref{ss_costmodel}). Finally, the best NN dataflow combination with minimal cost can be selected.   


\section{Experimental results}
This section demonstrates the strength of CMDS in terms of energy/latency/area. The dataflows found by CMDS with multi-bank-memory-based data reorganization will be compared to schedules found by a memory-unaware layer-wise scheduler under different BD/PD/MD data layouts. Two different baseline designs are used to benchmark the benefits brought by memory layout optimizations against memory-unaware scheduling: a.) an accelerator without a flexible memory interface, i.e., it is not equipped with a data reshuffling buffer in hardware. Supporting multiple different SUs hence comes with repeated memory accesses in case of non-matching data layouts between any two data-dependent layers; b.) an accelerator with a reshuffling buffer, which can reorder outputs to the desired layout at the expense of additional area overheads. The area cost of the reshuffling buffer will also be assessed in this paper.
However, comparing the energy/latency/area of all 3 cases (CMDS-driven dataflows v.s. memory-unaware dataflows without data reshuffling hardware module v.s. memory-unaware dataflows with reshuffling buffer) requires a more fine-grained memory access cost model than present in current layer-wise schedulers. Such a model will first be introduced in subsection V-A, followed by the experimental results in subsection V-B.

\subsection{Cost model}
\label{ss_costmodel}
\textbf{PD correction parameter}: Assessing layer-wise dataflow optimizers, all of them rely on a similar memory access cost model: 1) The PD of a memory instance is a user-defined input. 2) The data access energy cost per memory access is a user-defined fixed cost, regardless of the data layout. Data-layout-unaware cost models \cite{timeloop, zigzag, maestro}
assume the memory ports can always be fully utilized and can get PD-sized data per access. However, the wasted memory accesses due to layout mismatches between data stored in memory and data needed in the PE array have to be taken into account, resulting in an under-utilization of PD. 
Following the general memory architecture of Fig. 3, under a given SU, the resulting PD under-utilization falls apart into two aspects: 1) the BD data layout mismatch, causing a subset of words in BD to be useful per memory access; 2) the MD data layout mismatch, causing data across fewer bank to be useful per memory access. 
\mv{For} the first aspect, accessing \mv{only} a subset of the words of the BD, 
Eq.~(\ref{eq2}) describes the effective number of words ($\#Word_{eff}$) that are accessed jointly from a single bank \mv{given a particular BD layout and SU-compatible PD}: 
\begin{equation}
\label{eq2}
\begin{aligned}
\# Word_{eff} = \prod_{F}^{}min(BD[F], PD[F])
\end{aligned}
\vspace{-0.1cm}
\end{equation}
with $F$ the SU factors ${OXu, OYu, Ku/Cu}$. $BD[F]$, $PD[F]$ indicate the unrolling factors along the BD, and PD, respectively. The $min(...)$ computes how many words can be accessed in parallel from one bank row under a particular BD and PD data layout. \mv{Unlike the dataflows stemming from the datalayout-unaware schedulers} CMDS-driven dataflows force the PD data layout to a superset of the BD layout, which can always access the full memory bank row.  

The second aspect considers the PD under-utilization when the MD layout across banks does not cover any unrolling factor of PD (for a given SU) beyond the BD unrolling factors, hence preventing \mv{to jointly} access data from multiple banks. The effective number of banks accessed in parallel ($\# Bank_{eff}$) is:
\vspace{-0.05cm}
\begin{equation}
\label{eq3}
\begin{aligned}
\# Bank_{eff} = min[\frac{PD}{BD},~\prod_{F}^{}min(\frac{MD[F]}{BD[F]},~ \frac{PD[F]}{BD[F]})] \\
\end{aligned}
\vspace{-0.05cm}
\end{equation}

The $\prod_{}^{}min(...)$ in Eq.~\ref{eq3} computes the potential number of parallel accessed banks based on a specific data layout of BD, PD, and MD, respectively. This number can not exceed the number of banks fitting the port width: $\frac{PD}{BD}$. The total practical data volume per memory access is then derived by the product of $Word_{eff}$ and $Bank_{eff}$. This allows computing a PD correction parameter ($PD_{eff}$), which implies the effectiveness of PD per memory access with data layout awareness:

\begin{equation}
\label{eq4}
\begin{aligned}
PD_{eff} = \frac{\# Word_{eff} \times \# Bank_{eff} }{PD}\\
\end{aligned}
\vspace{-0.05cm}
\end{equation}

The $PD_{eff}$ allows us to perform memory-aware cross-layer dataflow evaluations while leveraging any traditional layer-wise cost model \cite{timeloop, zigzag, maestro}. \mv{This is possible by incorporating} the latency/energy losses due to the PD under-utilization \mv{into the} PD parameter of the activation memory, \mv{present in any} cost model. \mv{By replacing $PD$ as} $PD_{adjust}= PD_{eff} \times PD$, leaving all other settings untouched, \mv{all energy and latency losses due to layout mismatches are automatically accounted for.} 
The applied flow is summarised as follows: First, let the layer-wise schedulers, resp. CMDS find their optimal dataflow per layer with the physical memory port width $PD$. From these dataflows, derive the resulting $PD_{adjust}$.
Finally, rerun the cost models with the derived dataflows and the $PD_{adjust}$ to evaluate the energy and latency of the best dataflow from memory-aware CMDS and -unaware tools for each layer. 

\textbf{Reshuffling buffer cost model}: 
CMDS utilizes the existing multi-bank memory structure of activation memory to achieve efficient data reshuffling, only demanding $\frac{MD}{BD} \times \frac{PD}{BD}$ multiplexers to fetch/store PD-sized data from/to different banks. 
However, SOTA DNN accelerators with flexible spatial unrolling capabilities mainly exploit reshuffling buffers to reshape data from one layer to another. For comparison reasons, an area model is proposed to estimate the area overhead stemming from a reshuffling buffer (\#Reg, ignoring multiplexer cost) between any two data-dependent layers. ($layer_i$, $layer_{j}$):
\begin{equation}
\label{eq5}
\begin{aligned}
\# Reg =\prod_{F}^{}lcm(SU_i[F], ~RPD_{j}[F])
\end{aligned}
\vspace{-0.1cm}
\end{equation}
The $\prod_{}^{}lcm(...)$ in Eq.~\ref{eq5} computes the least common multiple (lcm) between the SU factors of a SU of $layer_i$ ($SU_i[F]$) and the unrolling factor of the read port of $layer_{j}$ (i.e., $RPD_{j}[F]$ of $layer_{j}$'s SU). The calculation obtains the number of $layer_i$'s final outputs that must be buffered to reorganize the data towards the RPD data layout of $layer_{j}$'s SU. 
The largest $\# Reg$ required by two data-dependent layers in the NN determines the size of the final reshuffling buffer for the complete NN.
Sacrificing area, a reshuffling buffer can avoid massive inefficient memory accesses. 

\vspace{-0.05cm}
\begin{table}[htbp]
\caption{Accelerator hardware configuration}
\vspace{-0.4cm}
\begin{center}
\begin{tabular}{|c|c|c|c|c|c|}
\hline
\textbf{Accel-}&\multicolumn{5}{|c|}{\textbf{Hardware Configuration}} \\
\cline{2-6} 
\textbf{erator} & \textbf{\textit{PE nb.}}&  \textbf{\textit{BD[b]}}&  \textbf{\textit{PD[b]}}&  \textbf{\textit{MD[b]}}&  \textbf{\textit{Mem Size[kB]}}   \\
\hline 
    ISSCC'22 \cite{diana} & 256 & 128 & 128 & 4096 &256  \\
    \hline
    VLSI'21 \cite{koen}  & 2048 & 128 & 1024 & 2048 &1024  \\
    \hline
    Proposed Archi.  & 1024 & 64 & 128 &  1024 & 512 \\
   \hline
\end{tabular}
\vspace{-0.5cm}
\label{tab1}
\end{center}
\end{table}
\label{ss:exp}

\subsection {Benchmarking and SOTA comparison}
To validate the benefits of CMDS, the execution latency and energy of mapping four NNs (ResNet20, ResNet18, DarkNet53 \cite{yolo}, and MobileNetv2) are evaluated separately for the 3 cases mentioned at the beginning of Section V. 
\mv{This benchmarking is moreover repeated for} three different hardware templates, whose configurations of interest are listed in Table~\ref{tab1}. In addition to \mv{published architectures}, \cite{diana} and \cite{koen}, we also propose a new architecture with a smaller BD, large MD, and $PD<MD$, supporting higher flexibility in memory data patterns to assess the impact on latency/energy due to different memory layouts. 
For a fair comparison, the cost estimations of three templates are normalized to 16nm FinFET CMOS technology. Although CMDS is compatible with any layer-wise cost estimation framework, the results reported in this subsection use the ZigZag framework. 

\begin{figure}[tb]
\centering 
\includegraphics[width=1\columnwidth]{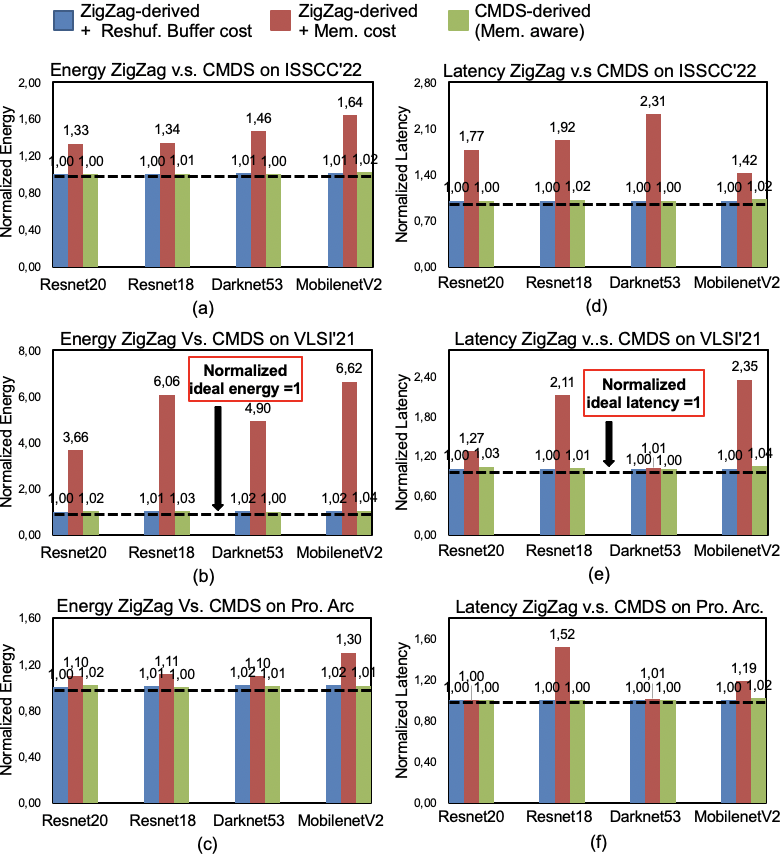}
\caption{Comparison between ZigZag and CMDS}
\vspace{-0.5cm}
\label{fig:exp} 
\end{figure}

\textbf{Energy analysis}: Fig.~\ref{fig:exp} (a)-(c) show for each hardware template the energy cost of executing the benchmark NN with the memory-unaware optimized dataflows and memory-aware CMDS.
All cases are normalized to the ideal memory-unaware energy without any data layout mismatch cost. The wasted memory accesses cost for the dataflows, and the cost of reshuffling buffer are derived as explained in Section \ref{ss_costmodel}.
These results indicate that the CMDS-driven dataflow brings less than $4\%$ energy degradation compared to the ideal memory-unaware end-to-end inference execution energy across all benchmarks and architectures, in sharp contrast to the naive layer-wise memory-unaware execution schedules of existing SotA scheduling frameworks. The reason is that instead of naively choosing the optimal SU, CMDS exploits one of the near-optimal SU alternatives to avoid data layout mismatches and, as such, come to a network-wide energy optimum. For example, when running MobileNetV2 on VLSI'21 \cite{koen}, around $5.6\times$ energy overhead is caused by inefficient memory accesses when simply combining the most energy-optimal SU of each layer derived by ZigZag. This stems from the mismatch between the design's BD (=16) and the many MobileNetv2 layers with OX and OY dimensions which are not multiples of 16. A memory-unaware framework favors lower $OX_u$ and $OY_u$ unrolling factors for high PE utilization, but this results in poor memory utilization.   
Finally, notice that all networks run more efficiently when executed on the proposed architecture, characterized by a $PD<MD$ with a small BD. This architecture brings more flexibility to its memory access schemes, as the ports can read from a flexible subset of the banks, resulting in fewer ineffectual memory accesses. The experimental results show that using the reshuffling buffer can avoid wasted memory accesses and only bring small energy costs because of the low energy consumption of register access. However, the overhead of reshuffling buffer comes at a considerable area cost which will be analyzed later. 
Finally, notice that all NN run more efficiently when executed on the proposed architecture, characterized by a $PD<MD$ with a small BD. This architecture brings more flexibility to its memory access schemes, as the ports can read from a flexible subset of the banks, making fewer wasted memory accesses. 

\textbf{Latency analysis}: In Fig.~\ref{fig:exp} (d)-(f), the end-to-end normalized latency is shown for each case. ISSCC'22 \cite{diana} benefits most from CMDS in all cases. Specifically, the BD data layout of ISSCC'22 is grouped by 16 OXs in their design. At the same time, the optimal SU derived by the memory-unaware scheduler always has smaller "$OX_u$" for most layers of Darknet53, resulting in a massive amount of partial BD reads/writes. We notice that some SUs with $OX_u=16$ match perfectly with the BD data layout of \cite{diana}, achieving the same performance as the layer-independent-optimized optimal SU. Nevertheless, the memory-unaware scheduler randomly chooses one of its optimal SU when multiple SUs have the same estimated cost, while they can be vastly different in memory access efficiency. Fig.~\ref{fig:exp}(f) shows that the proposed architecture with a smaller BD performs well. The smaller bank size of the proposed architecture enables full BD read/write. Besides, CMDS further exploits the flexibility provided by large MD, allowing to fully utilize PD and avoiding stalls. Notice that the reshuffling buffer latency costs are ignored in these results because the latency overhead of this operation can generally be hidden underneath the compute cycles. Nevertheless, the reshuffling buffer brings a high design complexity, demanding different data reorganization schemes to be compatible with diverse SUs.

\begin{table}
\caption{Area cost of reshuffling buffer}
\label{table:reg}
\centering
\begin{tabular}{|c|c|c|c|}
\hline
Accelerator & ISSCC'22 \cite{diana} & VLSI'21\cite{koen} &  Prop. Arc. \\
\hline
Model&\multicolumn{3}{|c|}{{Register number / 8b}} \\
\cline{2-4} 
\hline
ResNet20 &  2048 & 16384 & 8192 \\  \hline
ResNet18 &  4096 & 32768 & 16384  \\ \hline
DarkNet53  &  2048 & 32768 & 32768  \\ \hline
MobileNetV2  & 8192 & 65536 & 32768  \\ \hline
\end{tabular}
\end{table}

\begin{figure}[tb]
\centering 
\includegraphics[width=0.65\columnwidth]{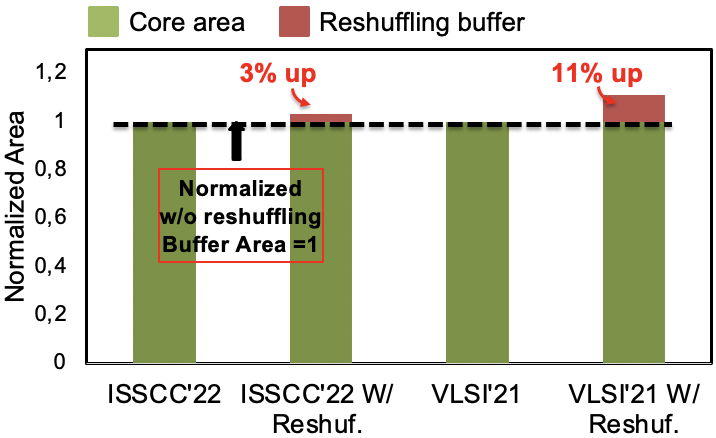}
\caption{Accelerator area W/O and W/ reshuffling buffer} 
\vspace{-0.6cm}
\label{fig:area} 
\end{figure} 

\textbf{Area analysis}: The reshuffling buffer \mv{enables to} reshape data \mv{when} switching \mv{from} one SU to another, with negligible energy and latency overhead, as shown in Fig.~\ref{fig:exp}. However, we can not underestimate the extra area overhead \mv{due to} the reshuffling buffer.
Table~\ref{table:reg} lists the \mv{required} \#Reg \mv{(Eqn.(5))} for \mv{the} reshuffling buffer \mv{in each of} the three hardware templates. 
\mv{This can be compared to} the total accelerator area of ISSCC'22 and VLSI'21, \mv{as done in Fig.~\ref{fig:area},} \mv{to assess the} significance of the reshuffling buffer area overhead on \mv{the} full system. 
The normalized area of \mv{the} accelerators is scaled based on 16 nm FinFET CMOS technology to be fairly compared. 
\mv{Results show that overhead is limited for ISSCC'22, yet} VLSI'21 \mv{would} needs 65536 8b registers for executing MobileNetV2 under the optimal layer-independent-optimized SU per layer, bringing a $11\%$ area increase. This is due to MobileNetV2's diverse layer topologies, resulting in distinct SU's for sequential layers. 
The bulky reshuffling buffer \mv{drastically reduces} area efficiency for memory-unaware schedul\mv{ed architectures}. In contrast, the proposed CMDS data reshaping \mv{requires only} a few multiplexers \mv{as} CMDS exploits the multi-banked structure. \mv{Moreover, the multi-bank structure is typically already natively present in designs using} large on-chip memories, \mv{typical for DNN accelerators, and hence comes without additional cost overhead.} 
We have noticed that the flexibility of CMDS is directly proportional to the number of memory banks. However, there is a tradeoff between the number of memory banks, the flexibility of CMDS, and the area required to implement the memory banks. In future work, we will investigate the optimal number of memory banks that balances these three parameters. We aim to find the number of memory banks that best balance flexibility, area, and the number of banks.

\section{Conclusion} 
Schedulers optimize the spatial and temporal dataflow to efficiently deploy DNNs with diverse layer topologies on CNN hardware accelerators. However, SotA DNN scheduling frameworks fail to properly take the memory constraints into account to \mv{jointly} optimize the data layout and the execution schedule.
Therefore, this work proposes CMDS, a framework enabling cross-layer memory-aware dataflow optimization to minimize execution energy/latency for end-to-end NN inference. 
CMDS can exploit the parallelism of multi-banked memories to achieve low-cost data reshuffling with increased dataflow flexibility. CMDS introduces the concepts of Bank width (BD), Port width (PD) as well as Memory width (MD) to represent the memory layout compactly, enabling to co-optimize the data layout together with the dataflow options to achieve end-to-end inference energy/latency minimization. Compared with the state-of-the-art (SOTA), which performs layer-optimized memory-unaware scheduling, CMDS achieves up to $5.5\times$ energy reduction and $1.35\times$ latency reduction with negligible hardware cost.  

\section{Acknowledgment}
This project has been partly funded by Vlaio IHearU project, the European Research Council (ERC) under grant agreement No. 101088865, the European Union’s Horizon 2020 programme under grant agreement No. 101070374, the Flanders AI Research Program and KU Leuven.

\end{document}